\documentclass{article}[12pt,aps,prb,preprint]
\usepackage{amssymb,amsmath,amsthm}
\def\la{\langle}
\def\ra{\rangle}
\def\ot{\otimes}

\begin{document}

\title{ Entangled states : Classical versus Quantum} 

\author{S. Kanmani, \\ Materials Physics Division, IGCAR, \\ Kalpakkam, India.\\ kanmani@igcar.gov.in \\}

\maketitle

\abstract{ Quantum mechanics of composite systems, gives rise to certain special states called entangled
states. A physical system, that is in  an entangled state displays an intricate  correlation between its
subsystems. There are also some composite  quantum states  ( classically correlated states or separable states
) that are not entangled. It is generally claimed, often without a rigorous proof to support,  that these
intricate correlations  of an entangled state cannot occur in a classical system.  This  expository article,
provides an elementary proof that entangled states cannot arise in the setting of classical mechanics. In
addition,  a detailed description of the origin of entanglement in  quantum systems is  included. The 
mathematical concepts that are necessary for this purpose are presented. The absence of entanglement in the
classical setting  is due to the fact that  every pure classical state of a composite system is a product
state, that is, a tensor product of two pure states of the subsystems. In contrast, there are pure composite
quantum states that  cannot   be expressed in the form of a product state or even by a convex sum of product
states. Roughly speaking, this is because classical states are positive valued functions on the phase-space
while quantum states are positive linear operators. The structure of the tensor product between two
commutative spaces of  scalar valued functions  is drastically different from that of the tensor product
between two non-commutative spaces of linear operators. In other words, entanglement is a non-commutative
phenomenon. }

\begin{center}

\noindent
{\bf Contents} 

\end{center}
\noindent
1.0 Introduction\\
2.0 States in classical mechanics\\
2.1 Composite classical systems and their states\\
2.1.1 Product states and separable states\\
3.0 States in quantum mechanics\\
3.1 States as positive operators\\
4.0 Composite quantum systems and their states\\
4.1 Composite quantum systems\\
4.1.1 Linear functionals and dual vector spaces\\
4.1.2 Bilinear forms\\
4.1.3 Tensor product of vector spaces\\
4.2 States of composite quantum systems\\
4.2.1 Separable states and entangled states.\\
5.0 Appendix (A-E)\\

\begin{quotation}				 

In  my opinion, the mathematics of last hundred years did not produce anything comparable to quantum theory or
general relativity in terms of the resulting change of our total world perception. But I do believe that
without the mathematical language physicists could not even say what they were seeing. 

\hfill - Yuri. I.Manin

\end{quotation}

\section{Introduction} 
\noindent
The strategy  of decomposing a complex object into simpler parts pervades  science. Thus, one tries to
understand a quantum mechanical  state of a composite system, $^{1}$ comprising of two particles  in terms of
its constituents, the single particle states. In that context, there arise certain composite states, called
entangled states $^{1}$ in which  the  subsystems  display a remarkable correlation between them. For example,
knowing the  state of one of the particle the state of the other can be predicted with certainity. It is
generally said that entanglement is a quantum phenomenon,  there by  implying such states do not arise in the
context of classical mechanics. For example, the article $ ^{2} $ states, `` Entanglement is a peculiar
property of quantum world that has no classical analog, .. ". The aim of this article is to provide a
pedogogical introduction  that clarifies the above  statement. \\

\noindent  We start with classical mechanics in section 2.0, where the motivation for representing a state as
a probability density function on phase-space is given. Section 2.1 considers the cartesian product of phase-
spaces as a composite classical system and looks at  the  nature of product states and separable states. The
result that every classical composite state is a separable state and hence  is a non-entangled state is
obtained in 2.1.1. Section-3 and section-4 are devoted to quantum systems and states. Section 3.0 begins with
the notion of a pure state  as a  vector of unit norm and contains a detailed discussion of mixed states and
their mathematical representation. Section 3.1 introduces the notion of density matrices; positive operators
with unit trace. Section-4.1 is a self-contained, rigorous introduction to tensor products. Finally, section
4.2 investigates the nature of composite pure states and demonstrates that every quantum mechanical pure state
associated with a non-elementary tensor is an entangled state.\\

\noindent
Readers interested in quantum information theory and those who wish to go beyond the modest aim of this
article may refer to $^{3, 4 }$ for more details. \\

\section{States in classical mechanics}    
\noindent 

In classical mechanics, we represent a state of a particle by specifying  a point $ x_0$ in the relevant
phase-space $X$. Recall, a point in a phase-space encodes both position and momentum of the particle.
Equivalently, such a state could also be represented by a scalar valued function, $ f:X \rightarrow R$, such
that $f(x) $ is $1$ when $x =x_0 \in X$ and $f(x)=0$ for all $ x \ne x_0$. This function  $f$, can be
interpreted  as a probability density function defined on the phase-space $X$. Such a state is called a pure
state in the context of classical mechanics or classical statistical mechanics.$ ^{5} $ A generalisation of
this notion, is a  probablity density function $g$, defined on the phase-space $X$,  such that $g(x_k)= p_k >
0$, for a finite set of points  $ \{ x_k \in X : 1 \le k \le n \} $, such that the  sum $ \sum_{k=1}^n p_k =1
$. The rest of the points in $X$, naturally, are assigned the value of zero probability. Such a classical
state is called a mixed state.$^{5}$ Thus, in general a classical state is a probability density function
defined on a  phase-space.\\

\noindent

{\bf Remark: }   Mixed states model a situation in which we are not able specify  the state sharply by a
single point on the phase-space; but can only assure that the system could be in any one of a finite set 
points, whose probability assignment is non-zero. Observe, that  the  real system is actaully in  one of those
points. In others words, mixed states model our ignorance of the state of  the actual system. This is analgous
to the notion of mixed state in quantum mechanics.\\

\noindent
{\bf Definition-1 } A classical state $f$, associated with a physical system on a phase-space $X$, is a
probability density function on $X$. That is, a classical state $f : X
\rightarrow R $, is a positive valued function such that  $f(x) \ge 0 $ for every  $ x \in X $ and
 $ \int_{X} f dx = 1 $. \\

\noindent {\bf Note:} For the sake of mathematical simplicity, we shall consider only those states $f$, for
which the following set $\{ x \in X : f(x) \ne 0 \} $, called the support of $f$, is a finite set. If $ \{ x_k
\in X : 1 \le k \le n \} $ is the support of a state  $f$, then the integral $ \int_{X} f dx = 1 $, that
occurs in the above definition reduces to  the sum $ \sum_{  i=1 }^{n} f(x_i) =1 $.\\

\noindent

{\bf Definition-2 } A classical state $f_{x_{0}}$ on $ X $, is called a pure state, if the total probability
of unity is assigned to a single point $ x_0 \in X $. That is, $ f_{x_{0}} : X \rightarrow R $ such that

\[  f_{x_{0}}(x) = \left \{ \begin{array}{rll}
                1 & \mbox{if} & x =x_0  \\
		0 &  \mbox{if} & x \ne x_0 ;
		\end{array} \right.
\] \\

\noindent
In this way every point in the phase-space $ X $ gives rise to a pure state.\\

\noindent
What is the relation between pure states and mixed states ? We shall show  that 
every mixed state is generated, in a sense, by a set of pure states. First, we observe that the set of
all scalar valued functions on the phase-space $X$, is a vector space. Suppose, $f$ and $g$ are two scalar
valued functions on $X$. Then one can define their sum $(f+g)$, which is another function on $X$ as follows.
Thus $(f+g) : X \rightarrow R $, where $(f+g)(x):= f(x) +g(x) $, for every $x \in X $. Similarly, one can define
the multiplication of a scalar $\alpha \in R $ with $f$, as $(\alpha f) : X \rightarrow R,$ where
$(\alpha f)(x):= \alpha \times f(x) $, for every $x \in X  $. Treating these two operations as vector addition
and scalar multplication respectively, one verifies that the set of all scalar valued functions on $X$, becomes
a vector space. Clearly, classical states are  elements of this vector space. Next, we introduce the notion of
convex combinations of vectors.\\

\noindent
{\bf Definition-3} Let $S =  \{ v_i : 1 \le i \le k \} $, be a set of vectors. Then any vector of the form  $
\sum_{  i=1 }^{k} a_i v_i $, where $ 0 \le a_i \le 1 $ for  $1 \le i \le k $ and $ \sum_{  i=1 }^{k} a_i =1 $
is called a convex combination of vectors from $S$.\\  

\noindent
${\bf Examples: } $\\ 

\noindent
1) Let  $ S=\{ v_1,v_2 \}$, where $v_1$, $v_2$ are two distinct  vectors on the plane.
Then the set of all convex combinations of $v_1 $ and $v_2$ is the set $ \{ p v_1 +(1-p) v_2 : 0 \le p \le 1
\} $. Geometrically, this set is the line segment $ \overline{v_1v_2} $, with $v_1$ and $v_2$ as their end
points.\\ 

\noindent
2) Let $T= \{ v_1, v_2, v_3 \} $ be  a set of three non-collinear vectors on the plane. Then the set of
all convex combinations of $T$, is the set of all the  points of the triangular domain, whose vertices
are the points $v_1, v_2 $ and $v_3$. \\

\noindent
Now we are ready for the relation between pure and mixed states.\\

\noindent
{\bf Proposition-1 } 
\noindent
Every classical state is either a pure state or a convex combination of pure states. That is, every mixed
state is a convex combination of pure states. \\

\noindent
{\bf Proof:} By definition-1 a state $f$, on a phase-space $X$ is a probability density function on $X$.  By
our assumption, the support of $f$ is a finite subset of $X$. That is, $f(x_i)=p_i >0 $  for a finite subset
$\{ x_i : 1 \le i \le n \}$  of $X$, and $ \sum_{  i=1 }^{n} p_i =1 $. Such a function can be expressed as $ f
= \sum_{  i=1 }^{n} p_i f_{x_{i}} $, where $ f_{x_{i}} $, represent pure states, for $ 1 \le i \le n $.
Recall, the function  $ f_{x_{i}} $, is defined such that $ f_{x_{i}}(x)=1 $, when $x=x_i$ and $
f_{x_{i}}(x)=0 $ for every other $x \in X$. Then, $ f(x_k) = \sum_{  i=1 }^{n} p_i f_{x_{i}} (x_k) = \sum_{ 
i=1 }^{n} p_i \delta_{i k} = p_k $, where $ 1 \le k \le n $ and $\delta_{i k}= 1 $ if $ i=k $ and $ \delta_{i
k}= 0 $ if $ i \ne k $. Note, $f$ is a convex combination of pure states. If $n=1 $ then $f$ is a pure state.
Thus, by construction any state $f$, is either a pure state or a convex combination of pure states. A
probability density function which assigns a non-zero probability to two or more phase-space points is called
a mixed state.\\

\noindent

Later, in section-3.1, we shall show  that a quantum state is characterised by a positive linear operator with
unit trace,  called density operator. Observe, the similarities between classical and quantum states. 
Positive linear operators of qunatum mechanics correspond to positive scalar valued functions on phase-space
of classical mechanics. Similarly, the  condition of unit trace for a quantum state corresponds to the
condition of normalisation; a necessary condition for a positive valued function to be a probability
density.\\

\subsection{Composite classical systems and their states:} 

\noindent
Consider a  particle, called particle-1, whose phase-space is the set $X$. Similarly, let $Y$ be the 
phase-space of another particle, called particle-2. The collective  system of particle-1 and particle-2, put
together constitutes   a  composite classical system. The phase-space of this composite system is the
cartesian product of $X$ with $Y$, that is, the set $X \times Y  =\{(x,y) : x \in X, \, y \in Y \}.$ Clearly,
as discussed above, the states of this composite physical system are probability density functions on the set
$ X \times Y$. \\

\noindent
Since every composite state is either a pure state or a convex combination of pure states, we shall look at
the pure states first. Any  probability density function on $X \times Y$, whose total probability is assigned
to a single point,  $(x_{0},y_{0}) \in X \times Y $ is a composite pure state. Explicitly, $ h_{(x_{0},y_{0})} : X\times
Y \rightarrow R $ is a composite pure state, where \\

\noindent

\[ [h_{(x_{0},y_{0})}](x,y) = \left \{ \begin{array}{rll} 1 & \mbox{if} & (x,y) =(x_0 ,y_0)  \\ 0 & 
\mbox{if} & (x ,y ) \ne (x_0,y_0).  \end{array} \right. \] \\		 

\noindent
It is easily verified that  this is compatible with proposition-1. In other words, an arbitrary  composite
mixed state is the same thing as a convex combination of composite pure states of the above form.\\

\subsubsection{ Product states and separable states }
\noindent
What is the relation between  the pure states of $X\times Y$, the composite system, to the pure states of the
subsystems $ X $ and $Y$ ? Specifically, let $h_{(x_0,y_0)}$ be a composite pure state as defined above. Then
let  $f_{x{_0}}$ and $g_{y_{0}}$ be the pure states of the subsystems $X$ and $Y$ respectively. Explicitly, 
$ f_{x{_0}} : X\rightarrow R $, and $ g_{y_{0}} : Y \rightarrow R $,   are such that,

\[ f_{x_{0}}(x) = \left \{ \begin{array}{rll}
                1 & \mbox{if} & x =x_0   \\
		0 &  \mbox{if} & x \ne x_0 .
		\end{array} \right.
\]		and

\[ g_{y_{0}}(y) = \left \{ \begin{array}{rll}
                1 & \mbox{if} & y =y_0  \\
		0 &  \mbox{if} & y\ne y_0 . 
		\end{array} \right.
\]	\\

\noindent
Given two  functions  $ f :X\rightarrow R$, and $ g:Y\rightarrow R $  one can define $f\ot g $,
their tensor product as $ f\ot g : X\times Y \rightarrow R $, where $[f\ot g] (x,y) = f(x)\times g(y)$. In the
last equality, the product on the right hand side is the product of the real numbers $f(x)$ and $g(y)$.
Roughly, this is like multiplying, $P(x)$, a polynomial in the variable $x$, with $Q(y)$, another polynomial
in the variable $y$, to get $R(x,y)= P(x) \times Q(y) $, a polynomial in the variables $x$ and $y$.
Essentially, for the space of scalar valued functions, tensor product is the same as the - natural-
multiplication of functions as indicated above.\\

\noindent
Thus, the tensor product of  pure states of the subsystems $f_{x_{0}} $and  $
g_{y_{0}} $ is of the form $ f_{x_{0}}\ot g_{y_{0}}= f_{x_{0}}\times g_{y_{0}}$. Clearly, $[f_{x_{0}}\times
g_{y_{0}}] (x,y)=f_{x_{0}}(x)\times g_{y_{0}}(y)= \delta_{x_0 x} \times \delta_{y_0 y}.$ Hence, this product
of two functions takes the value of $1$ if and only if $x=x_0$ and $y=y_0 $ and takes the value of $0$ at all
other points. Explicitly, \\

\noindent

$ f_{x_{0}}\ot g_{y_{0}} : X\times Y \rightarrow R,$ such that 

\[ [f_{x_{0}}\ot g_{y_{0}} ](x ,y)=f_{x_{0}}(x) \times g_{y_{0}}(y)        = \left \{ \begin{array}{rll}
                                                        1 & \mbox{if} & (x ,y ) =(x_0,y_0)  \\
		                                        0 &  \mbox{if} & (x ,y) \ne (x_0,y_0). 
		                                        \end{array} \right.
\] \\
 
\noindent  
Note that this is exactly the same as the pure state $ h_{(x_{0},y_{0})}$, of the composite system $ X \times
Y $. Thus, $ f_{x_{0}}\ot g_{y_{0}} = h_{(x_{0},y_{0})} $. In other words, every pure state of a classical
composite system is in the form of a product of pure states of the subsystems. The composite states of the
form $ f_{x_{0}}\ot g_{y_{0}} $ are called {\bf product states }. \\

 \noindent
{\bf Definition-4 } A  composite state of the form $f \ot g $, where  $f$ and $g$ are the states of the
subsystem is called a {\bf product state }.\\ \noindent Thus we have proved the following proposition.\\

 \noindent
 {\bf Proposition-2} Every classical composite pure state is a tensor product of pure states of the
 subsystems. Thus, every  pure state of a composite classical system is a product state.\\

\noindent
{\bf Note :} This is not true for a composite quantum system. In other words, as we shall see, there are pure
states in a composite quantum system which cannot be expressed in the form of a product state. In fact, they
cannot be even written in the form of a convex combination of product states.\\ 

\noindent
{\bf Definition-5 } A  composite state of the form $\sum_{i}^n p_i f_i \ot g_i $, where  $\{ f_i \} $ and $\{
g_i \} $ are the states of the subsystems is called a separable  state. Here, $ 0 \le p_i \le 1 $ for  $1 \le
i \le n $, and $ \sum_{  i=1 }^{n} p_i =1 $. If $n=1 $ this becomes a product state. Thus, a separable state
is either a product state or a convex combination of product states.\\

\noindent
{\bf Definition-6 } A composite state that is {\bf not } a separable  state is called an entangled state.
 Thus, any state that cannot be expressed as a convex combination of product states is an entangled
state. \\ 

\noindent 
By proposition-1, every state is either a pure state or a convex combination of pure states. In the
case of a composite classical system, every pure state is a product state (cf. Proposition-2). Thus, every
classical composite state is either a product state or a convex combination of product states. Hence, by the
definition-5 of separable states, every classical composite state is a separable state. Thus we have the
following proposition.\\

\noindent
{\bf Proposition-3 } Every classical composite state is a separable state. Equivalently, there are no
entangled states in a classical composite system.\\ 

\noindent
{\bf Remark: } Given a  composite classical state $h=h(x,y)$, on $X \times Y $ one can associate a state
$g(y)$, of the subsystem  $Y$ in a natural way. This is done by partially integrating the state $ h(x,y)$, the
probability density, with respect to  the variable $ x$,  resulting in a marginal probability density $g(y)$
in $Y$. It is easily verified, that  every classical composite pure state thus reduces to a pure state of a
subsystem. That is, $ \int_{X} h_{( x_{0},y_{0})}(x,y) dx = \int_{X} f_{x_{0}}(x) \times  g_{y_{0}}(y) dx
=g_{y_{0}}(y) $, where, the states are pure states as defined above.  This is not true for a quantum system,
where a partial trace $^{1}$ of a pure composite state may result in a mixed sate of the subsystem. This was
first observed by schroedinger. Partial tracing is the quantum analog of partially integrating a composite
state over one of the variables of the subsystems.

\section{States in quantum mechanics} 
\noindent
We shall assume that all our vector spaces are  finite dimensional complex vector spaces. Recall, that the
quantum mechanical observables associated with position and momentum  cannot be modelled $^{6}$ on a
finite dimensional vector space. For example, in the context of an electron, only its spin degree of freedom
can be modelled on a finite dimensional vector space. \\

\noindent
A pure state of a quantum mechanical system is characterised  by a vector $x$ of unit norm in a 
Hilbert space H.  As is well known, physical observables are  represented by self-adjoint operators acting on
that Hilbert space. The expectation value of an observable $A$, when the system is in a pure state $x$  is
given as $ \la x, A x \ra $. Here, $ \la u, v \ra $ denotes the inner product between the vectors $u$ and $v$
of the space $H$. We shall adopt the convention in which $ \la x, \alpha y \ra $ = $ \alpha \la x, A x \ra $
and  $ \la \alpha x,  y \ra $ = $ \overline{\alpha} \la x, y \ra $, where $\overline{\alpha}$ denotes the
complex conjugate of the complex number $\alpha $.\\

\noindent 
Intuitively, a mixed state is a probability density  defined on a set of pure states. A simple
example of a mixed state is a set containing  two  pure states $\{ x_1, x_2 \} $,  such that the state $x_1$
is assigned a probability of $p_1$ and the state $x_2$ is assigned the probability $p_2= 1-p_1$. Though, the
actual system is strictly in only  one of those  two  pure states, we do not know which one of $ \{ x_1, x_2
\} $ is that. Hence, we model this state of uncertainity through a probability distribution  on the set of
possible pure states. Until we find an appropriate mathematical representation for a general mixed state, we
shall denote this mixed state  as $S_m=\{ (x_1,p_1), (x_2,p_2) \} $; a set of ordered pairs, whose first
component is a pure state and the second component is the probability associated with it. The expectation
value of an  observable $A$, when the system is in the  mixed state  $S_m$, has to be the weighted sum of $
\la x_1, A x_1 \ra $ and   $ \la x_2, A x_2 \ra $, with their respective probabilities $p_1$ and $p_2$ as
weights. Thus, the expectation value of an observable $A$, in the mixed state $S_m$  is $p_1 \la x_1, A x_1
\ra + p_2  \la x_2, A x_2 \ra $, where $p_1 +p_2 =1 $. It is important to understand that a mixed state can
not be represented as a vector in $H$. Suppose we try to represent the mixed state $S_m$, as  a vector $ x=
p_1 x_1 + p_2 x_2 $, where $p_1 + p_2 =1$; then the expectation value of an observable $A$, in the state
$S_m$  is  $ \la x, A x \ra =  \la p_1 x_1 + p_2 x_2 , A (p_1 x_1 + p_2 x_2) \ra = p{_1}^2 \la  x_1, A x_1 \ra
+  p_1p_2 \la x_1, Ax_2 \ra + p_1 p_2 \la  x_2, Ax_1 \ra + p_{2}^2\la x_2, A x_2 \ra $ = $ p{_1}^2 \la  x_1, A
x_1 \ra +2 p_1p_2 Re( \la x_1, A x_2 \ra ) + p_{2}^2\la x_2, A x_2 \ra $. In the above expression we have made
use of the fact that $A$ is self-adjoint and that  $ \la u, v \ra + \la v, u \ra $ is equal to two times the
real part (denoted as Re ) of the complex number $\la u, v \ra$. It can be verified that $ \la x, A x \ra $ as
defined by the expression above is not equal to $p_1 \la x_1, A x_1 \ra + p_2  \la x_2, A x_2 \ra $, the
correct expectation value of an observable $A$ in the state $S_m$.  This demonstrates that it is not possible 
to represent a mixed state   as a linear supersposition of pure state vectors.\\

\noindent
Hence, our aim  is to obtain a  mathematical representation of a mixed state that will satisfy the following
two conditions. i) Expectation value of  an observable $A$, in the state  $S_m= \{ (x_1,p_1), (x_2,p_2) \}$,
should be $p_1 \la x_1, A x_1 \ra + p_2  \la x_2, A x_2 \ra $. ii) Every mixed state should be  a convex
combination of pure states.\\

\subsection{ States as positive operators }

\noindent 
This aim is achieved by representing both pure and mixed states as a particular class of linear
operators acting on the Hilbert space $H$. Suppose $S$ is such an operator representing a quantum state, then
the expectation value of an observable $A$, in the state $S$  is now defined as $ Tr(AS) $, where $Tr(B)$
denotes the trace of an operator $B$. In such a generalization, a  pure state $x \in H $  is  represented as a
linear  operator $ P_x : H \rightarrow H $, defined by its action on $ u \in H$  as $ P_x(u) = \la x, u \ra x
$.  Then the expectation value of an observable $A$, in the state $P_x$ is $ Tr(AP_x)$. Now we prove that $
Tr(AP_x)= \la x, Ax \ra $ for any pure state $x $  and any observable $A$ as it should be. By definition,
trace$^{7}$ of a linear operator $T$  is defined as  Tr$(T)= \sum_ {i=1}^{n}  \la e_i, T e_i \ra $, where $
\{  e_{i}: 1 \le i \le n \} $ is any orthonormal basis of $H$. Given a $x \in H$, it is always possible to
find  an orthonormal basis $ \{  e_{i}: 1 \le i \le n \} $, of $H$ in which $e_1=x$. Then $Tr(AP_x) =  \la
e_1, (AP_x )e_1 \ra +\sum_ {i=2}^{n}  \la e_i, (AP_x )e_i \ra =  \la x , (AP_x) x \ra +\sum_ {i=2}^{n}  \la
e_i, (AP_x) e_i \ra = \la x, A x \ra $. This is because $P_x(x)= \la x, x \ra x = || x||^2 x = x$ and $
P_x(e_i)= 0 $, for every $ 2 \le i \le n $, by our  choice of orthonormal basis. \\

\noindent
By representing pure states $x_i$ as  $P_{{x_i}}$, the mixed state $ S_m $, could now be expressed 
as $ \rho = p_1 P_{x_{1}} +p_2 P_{x_{2}} $,  as a convex combination of pure states. Then, the expectation
value of an observable $A$, in the mixed state $S_m$ gets reproduced correctly as  $Tr (A\rho) = Tr[ A(p_1
P_{x_{1}} +p_2 P_{x_{2}})] = p_1 Tr(AP_{x{_1}}) + p_2 Tr(AP_{x_{2}}) = p_1  \la x_1, A x_1 \ra + p_2  \la x_2,
A x_2 \ra $. Here, we have used  the facts that $ Tr(A+ B) = Tr(A) + Tr(B) , \, Tr (\alpha A) = \alpha Tr(A)$ 
and the identity  $\la x ,Ax \ra = Tr(AP_x) $ that we have proved earlier.  Thus, we have obtained  a
mathematical representation of mixed states that is consistent with the two conditions stated above.\\

\noindent
Now, we shall show that $P_x$ can be characterised as  a self-adjoint, projection operator of rank one. First 
we shall introduce the notion of rank of a linear operator and show that the pure state $P_x$ is a rank one
linear operator. A linear operator is a mapping $ T: H \rightarrow H $, such that $ T (\alpha u + \beta v )=
\alpha T(u) + \beta T(v) $, for every $u, v \in H $ and every $\alpha, \beta \in C $. The range of a linear
operator $ T $, denoted as range(T) is the set $ \{ T(x) : x \in H \}$.  This set range(T), for any linear
operator $T$,  is  a  subspace$^7$ of $H$.  The rank of a linear operator $T$, is  by definition, the
dimension of  the  range $(T)$. When a linear operator is  represented by a matrix, its range is the
span of its coulumn (or equivalently row) vectors. Thus, the rank of a matrix $M$, is the maximal number of
linearly independent columns (or equivalently rows) of $M$.\\

\noindent
Recall, that the linear operator $ P_x : H \rightarrow H $, that represents a pure state  acts on a arbitrary
$u \in H$ in the following way. $ P_x  (u) = \la x, u \ra  x = z x$, where $\la x, u \ra $ denotes  the inner
product of vector $x$ with $u$ and hence is equal to a complex number $z$. Thus, $P_x$ maps any vector $u \in
H$ into  the one dimensional subspace spanned by $x$. Hence, $P_x$ is a projection operator and  the range of
$P_x$ is a one dimensional subspace of $H$. Thus, rank of $P_x$ is one.   Since $P_x(u)= \la x,u \ra x $ and
$P_x(v)= \la x,v \ra x $, it follows that $P_x$ is self-adjoint as $ \la v, P_xu \ra = \la v, \la x, u \ra x 
\ra = \la x, u \ra \la v, x \ra = \la \overline{\la v,x \ra} x, u \ra =  \la \la x,v\ra x,u \ra = \la P_xv,u
\ra $. Similarly, it follows that $P_x P_x = P_x $, because $ P_x (P_x(u))= \la x, P_x(u) \ra x = \la x, \la
x,u \ra x \ra x = \la x,u \ra \la x,x \ra x = \la x,u \ra x = P_x(u) $ for every $u \in H$. In Dirac's
notation $P_x$ is written as $ |x\ra\la x| $. We prefer $P_x$ over  Dirac's $ |x\ra\la x| $ as it is 
convenient in the context of tensor products ( cf. appendix-E for more on Dirac's notation ). Thus we have a
formal definition of quantum states as given below.\\

\noindent
{\bf Definition 7 } A pure state of a quantum mechanical system  modelled on  a Hilbert space $H$, is  a
self-adjoint, rank one  projection operator. We shall denote them as  $P_x$, where $x \in H $ and is of unit
norm.\\

\noindent
{\bf Definition 8} A mixed state of  a quantum mechanical system  modelled on  a Hilbert space $H$ 
is  a  convex combination pure states. Thus, if $\rho $ is a mixed state then $ \rho = \sum_{i=1}^k p_i
P_{x_{i}} $, where  $ \sum_{i=1}^k p_i =1 $ and $ P_{x_{i}} $, are  pure states for  $ 1 \le i \le k $.

\noindent   A  classical state is a probability density function and hence is positive valued. We shall show,
in a  sense, the  operators that represent quantum states also have certain positivity property just like the
classical states.\\

\noindent
Linear operators or equivalently matrices can be thought of as a  generalization of complex numbers. Suppose,
$T : C \rightarrow C $ is a linear operator acting on the one dimensional complex vector space $C$. Then, its
action on $ z \in C$ is as $T(z)= w_T z $, where $w_T$ is a fixed complex number. Equivalently, the $ 1 \times
1 $  matrix representation of $T$ is the complex number $w_T$. Then $T^*$, the adjoint of $T$  is represented
by $\overline{w_T }$, the complex conjugate of $w_T$. Thus the notion of adjoint is a generalisation of
complex conjugation. If $T$ is self-adjoint, then  $T =T^* $ or equivalently  $\overline{w_T } $ = $ w_T $.
This implies  that a  self-adjoint operator $T$ is represented by a real number $w_T$. Hence, self-adjoint
operators are like  real numbers. To summarise, if one  thinks of an arbitrary linear operator as a
generalized complex number, then self-adjoint operators are like  generalised real numbers.\\

\noindent
A pure quantum state $P_x$, being a self-adjoint operator is like a real number. Pushing this analogy between
operators and complex numbers further, we claim that $P_x$ is in fact like a positive real number. A complex
number  $z$ is a positive real number if and only if $z= \overline{w} w $ for some complex number $w$. Since,
adjoint is the appropriate generalisation of complex conjugation, we shall call an operator $T$ to be a
positive operator if $ T= B^* B$ for some operator $B$. \\

\noindent
{\bf Definition 9 }  An operator $ T: H \rightarrow H $ is called a positive operator if $ T=B^*B $ for some
operator $B$. Here $B^*$ denotes the adjoint of $B$.\\

\noindent
It is seen immediately that $P_x$ is a positive operator, because $ P_x^*  P_x = P_x  P_x =P_x $. As observed
earlier $ P_x P_x =P_x $ and $ P_x^*= P_x $ as $P_x$ is self-adjoint. Recall an operator  $T$, acting on a
Hilbert space is called self-adjoint if $ \la T u, v \ra = \la u, T v \ra $ for every $u,v \in H$. In the case
of complex vector spaces,  there is an another   definition for self-adjoint operators that is equivalent to
this. \\

\noindent
{\bf Proposition-4} If $H$  is a complex vector space then $T: H \rightarrow H $ is a self-adjoint operator if
and only if $ \la T u, u \ra = \la u, T u \ra  $  for every $u \in H$.\\ 

\noindent
{\bf Remark: } From the property of inner products $\la T u, u \ra $ is the complex conjugate of $ \la u, T u
\ra  $. Thus, in a complex vector space $H$, an operator $T$  is self-adjoint if and only  if $\la u, T u \ra
$ = $ \overline{ \la u, T u \ra }  $, or equivalently  if and only  if $ \la u, T u \ra $  is a real number
for every $ u \in H $.\\

\noindent
{\bf Proof: } (cf. Appendix-A ) \\ 

\noindent
Now  we  record an another definition of positive operators, which is equivalent to definition 9  in the
context of complex vector spaces. \\

\noindent
{\bf Definition 10} An operator $T: H \rightarrow H$, on a complex vector space $H$ is positive if  $ \la x, T
x \ra  \ge 0 $ for every $x \in H$. \\

\noindent
{\bf Proposition-5 } In a complex vector space $H$, the following two statements about a linear operator $T: H
\rightarrow H $ are equivalent.

\noindent  
1) $T = B^* B $ for some operator $B$. \\ 
2) $ \la x, T x \ra \ge 0 $ for every $x \in H$.\\

\noindent
{ \bf Proof     } ( cf. Appendix-B )\\

\noindent
{\bf Proposition-6}  A pure state of a quantum mechanical system $P_x$, is a positive, rank-one operator
of unit trace.\\

\noindent
{\bf Proof} : It has been shown earlier that $P_x $ is a rank-one linear operator. Now, we prove that $P_x$ is
a positive operator using definition-10. Since, $ \la u, P_x u \ra = \la u, \la x,u \ra x \ra = \la x,u \ra
\la u, x \ra = \la x,u \ra  \overline{\la x,u \ra } \ge 0 $, for any $ u \in H $, it follows that $P_x$ is a
positive operator. Here we have used the properties of inner product and the definition  of the linear
operator $P_x$, which acts on $ u \in H $ as  $ P_x(u) = \la x, u \ra x $.  Now we compute the trace of $P_x$.
By definition, Tr$(P_x)= \sum_{ i=1}^{n} \la u_i, P_x u_i \ra $, where $ \{  u_{i}: 1 \le i \le n \} $ is any
orthonormal basis of $H$. Choosing, an orthonormal basis of $H$, in which $u_1 =x $, one gets Tr$(P_x)= \la
u_1, P_x u_1 \ra  +\la u_2, P_x u_2 \ra + ....+\la u_n, P_x u_n \ra= \la x, P_x x \ra = \la x, \la  x,x \ra x
\ra = ||x||^4 =1$ as the later terms vanish and the norm of $x$ being one. Thus, $P_x$ is a positive,
rank one  operator with unit trace.\\

\noindent
Since, a general state is either a pure state or a mixed state, we have the following characterisation of  
a quantum state.\\

\noindent
{\bf Proposition-7 } A  quantum mechanical state   is a positive operator of unit trace. Such an operator
is   called   a density operator or matrix.\\ 

\noindent
{\bf Proof: }  A state is either a pure state or a convex combination of pure states. If it is a pure state
then by proposition-6 it is a positive operator of unit trace. A mixed state is a convex combination of pure
states.  Suppose $ \rho_1 $ and $ \rho_2 $ are two positive operators and  $ p_1 \rho_1 + p_2 \rho_2 $,  a
convex combination of them. Then, $ \la u , (p_1 \rho_1 + p_2 \rho_2) u \ra =  p_1 \la u, \rho_1 u \ra + p_2
\la  u , \rho_2 u \ra  \ge 0 $, as $ \rho_1 $ , $ \rho_2 $ are positive operators and $p_1, p_2 $ are positive
real numbers. Thus,  a convex combination of positive operators, is a positive operator.
Hence, a mixed state is a positive operator. Similarly, if tr$(A)=1 $ and tr$(B)=1$ then tr$ (p_1 A +p_2 B )$=
$p_1$ tr$A$+ $p_2$ tr$B$= $p_1+p_2=1 $. Thus it follows that a convex combination of unit trace operators is
an operator of unit trace. Since, pure states are  of  unit trace it follows that a mixed state, which is a
convex combination of pure states is  of unit trace as well.\\

\noindent

 Table-1   displays  the analogy between   classical states and quantum states.

\begin{table}
\caption{Analogy between classical and quantum states}
\begin{center}
\begin{tabular}{|lcr|}
\hline

  Property  &  Classical  &   Quantum \\
\hline
State  &  $f :X \rightarrow R $ &  $\rho : H \rightarrow H  $  \\ 

\hline
 Positivity  &  $f(x) \ge 0 $ ; $ x \in  X $  &  $ \la x, \rho x \ra \ge 0 $ ; $ x \in H $ \\
\hline
Normalisation & $ \int_{ X} f(x) dx =1 $ &  Tr $ \rho $ =1 \\
\hline
Pure state & $ \{ x \in X | f(x) \ne 0 \} $ - singleton set  & rank of $ \rho $ = 1  \\

\hline 

\end{tabular} 
\end{center}
\end{table}

\section{ Composite quantum systems and their states }
\noindent
A simple example of composite quantum system is a physical system that consists of two particles. For example,
a pair of electrons. The spin degree of freedom of a single electron is modelled on $\mathbb{C}^2$, a two
dimensional complex vector space. The composite object  of two electrons,  considering only the spin degree of
freedom,  is modelled on the vector space $\mathbb{C}^2 \ot \mathbb{C}^2 $, the tensor product space of  $
\mathbb{C}^2 $ with itself.  Hence, one should  consider the concept of  the tensor product of two vector
spaces. \\

\subsection{Composite quantum systems} 
\noindent
Now we begin our study of composite quantum systems. Before we define
notion of tensor product, we introduce the notions of linear functionals, the dual of a vector space and
bilinear functionals or forms.  

\subsubsection{Linear functionals and dual vector spaces}
\noindent
  
  Given a complex vector space $X$, consider a complex  valued linear mapping $\phi$, defined on $X$. That is,
$\phi : X \rightarrow \mathbb{C} $, such that $\phi(\alpha x_1 + \beta x_2 ) =\alpha \phi( x_1) + \beta \phi
(x_2)$ for every $ x_1, x_2 \in X $ and $ \alpha , \beta \in \mathbb{C} $. Such a $\phi $ is called a linear
form or a linear functional on $X$. For example, for a fixed $v \in X $, define a linear map $ \phi_v : X
\rightarrow \mathbb{C} $, where $\phi_v(x) := \la v, x\ra $.  It can be seen that, $\phi_v(\alpha x_1 +  \beta
x_2 )= \la v,\alpha x_1 + \beta x_2 \ra = \alpha \la v, x_1 \ra + \beta \la v, x_2 \ra = \alpha \phi_v(x_1) +
\beta \phi_v(x_2) $, for every $ x_1, x_2 \in X $ and $ \alpha , \beta \in \mathbb{C} $.\\

\noindent
Suppose  $\phi$ and $\psi $ are two linear functionals on $X$, then their sum $(\phi + \psi) $, is another
linear functional on $X$. This sum is defined  as ; $(\phi + \psi)(x) := \phi(x) + \psi(x)$, for every $x \in
X$. Similarly, the multiplication of a scalar $\alpha \in \mathbb{C}$ with a linear functional $\phi $ on $X$
results in a  linear functional denoted   as $ (\alpha \phi) $. This is defined as $ (\alpha \phi)(x) :=
\alpha \times \phi(x) $ for every $x \in X$. With these two operations, as one can verify, the set of all
linear functionals on $X$, becomes a vector space. This  is called the dual vector space of $X$ and is denoted
as $X^*$. Note, the zero element of this vector space $X^*$ is a linear functional $\phi_0$, such that $
\phi_0(x)= 0 \in \mathbb{C}$ for every $ x \in X$. Often, we shall denote the zero linear functional by $0$.
If $\phi$ is a non-zero linear functional, then there is a $ x \in   X$ such that $ \phi(x) \ne 0$. In
particular, if $\phi(x)= 0$ for every linear functional $ \phi \in X^*$, then $x=0$. Let $E= \{ e_1, e_2, ...
,e_n \} $ be  a basis of $X$. Then a linear map $T$  on $X$ gets  completely specified by the values  $ \{
T(e_k) :   e_k \in E \} $. For example, if $ \{e_1, e_2 \} $ is a basis of a two dimensional vector space $X$,
then there is a unique linear functional $ \phi \in X^*$, such that $ \phi(e_1)=1 $ and $\phi(e_2)=0 $. Later,
we shall make use of such   linear functionals.\\

\subsubsection{ Bilinear forms} 
\noindent 
We shall define a  tensor product space  as a dual vector space of the space of bilinear forms. Hence, we
shall introduce  the notion of a bilinear form. Suppose $X$ and $Y$ are two vector spaces. Then a complex 
valued function $ f$, defined on $ X \times Y$ is  called a bilinear form if it satisfies the following
conditions. 1)  $ f(\alpha x_1 + \beta x_2, y )= \alpha f( x_1,y ) + \beta f(x_2, y) $ for every $ x_1, x_2
\in X $,   $ y \in Y $ and $ \alpha , \beta \in \mathbb{C} $ and 2)  $ f(x,\alpha y_1 + \beta y_2)= \alpha f(
x,y_1 ) +  \beta f(x, y_2) $ for every $ y_1, y_2 \in Y $, $ x \in X $ and $ \alpha , \beta \in \mathbb{C }$.
That is, $ f $ is a function of two (vector) variables such that $f$ acts as a linear map in each variable
when the other variable is fixed. Now we shall look at an example of a bilinear form. Let $X$ be a vector
space and $X^*$ its dual. Then the map $b : X \times X^* \rightarrow \mathbb{C}$, where $ b(x, \phi ) :=
\phi(x) $, \, $ x \in X $, \, $ \phi \in X^* $, is a bilinear form.\\

\noindent
Note, that a bilinear form is a not a linear map. Clearly, the domain of a bilinear form, that is, the set
$X\times Y$ is not even a vector space. However, the  set of all bilinear forms from $ X \times Y$ to
$\mathbb{C}$, is a vector space. The sum of two bilinear forms and the multiplication of a complex scalar with
a bilinear form are defined  pointwise, just as we did in  the case of linear functionals. For example, if
$f$ and $g$ are two bilinear forms on $X \times Y $, then $(f+g)(x,y):=f(x,y)+ g(x,y) $, for every $(x,y) \in
X \times Y$. Similarly, one can define the multiplication of a complex scalar, with a bilinear form. We shall
denote this vector space, that is, the vector space of all bilinear forms on $X\times Y$ as $B(X\times Y)$.

\subsubsection{ Tensor product of vector spaces} 
\noindent
The notion of tensor product involves many abstract concepts. First of all, keep in mind that the symbol $
X\ot Y$, stands for a vector space. The  symbols $X$ and $Y$ in $ X\ot Y$, remind us that it has been created,
crudely speaking, by a sort of product or multiplication of two vector spaces $X$ and $Y$. The elements of $
X\ot Y$   are  vectors. However, to emphasize the fact that these elements were obtained  by the special
process of - tensor product - of two vector spaces, we shall call them tensors. The space $ X\ot Y$, contains
some elements that can be considered as if they were obtained by multiplying an element $ x \in X $ with 
another  element $ y \in Y $. We shall denote such an element as $ x \ot y $. Such elements are called
elementary tensors. Infact, every element in $X\ot Y$  is a sum of elementary tensors. Note, in the context of
the symbol $x \ot y $, that  $ x \in X $, $ y \in Y $ and  $  x \ot y \in X\ot Y $.\\

\noindent
Formally, the tensor product, $X\ot Y$, of the vector spaces $X$ and  $Y$ is  defined  as the dual space of
the vector space of bilinear forms $B(X \times Y) $. That is, if $ \tau \in X\ot Y $, then $\tau $ is a linear
functional from the vector space of $ B(X \times Y) $ to the space of complex numbers. Specifically, $ \tau:
B(X \times Y ) \rightarrow \mathbb{C}$, is defined such that $ \tau( \alpha f_1 + \beta f_2 )= \alpha
\tau(f_1) +\beta \tau(f_2) $, for every bilinear form $ f_1, f_2 \in  B(X \times Y)$ and  $\alpha, \beta \in
\mathbb{C} $. \\ 

\noindent
Hence, if   $ x \in X $ and  $ y \in Y $, then the symbol  $x \ot y$, as we defined above, denotes  a linear
functional on $  B(X \times Y) $. That is, $ x \ot y $ stands for a linear map from the vector space of $ B(X
\times Y)$ to the space of complex numbers. Formally, $  x \ot y :  B(X \times Y) \rightarrow \mathbb{C} $ and
the  action of  $x \ot y$ on a bilinear form $ f \in  B(X \times Y) $ is defined as ;  $ (x \ot y)
(f):=f(x,y)$.  If  $x' \ot y'$ is another  linear functional acting on $B(X,Y)$, then their sum denoted as $x'
\ot y' + x \ot y$ is defined as follows; $(x' \ot y' + x \ot y) f = (x' \ot y')f + (x \ot y) f = f(x',y') +
f(x,y)$ for every bilinear form $ f \in B(X\times Y)$. Similarly, the multiplication of a complex scalar
$\alpha $ with a linear functional results in another linear functional. This is done  by defining it as $
(\alpha (x \ot y))( f) := \alpha \times (x \ot y)(f)= \alpha \times f(x,y) $ for  every $\alpha \in \mathbb{C}
$. Thus,  $X\ot Y$, is the vector space of all linear functionals spanned  by the functionals of the form $x
\ot y $. Tensors of the form $x \ot y $, are called elementary tensors. Formally, $X\ot Y$ =span $ \{ x\ot y :
x \in X, y \in Y \}$.\\

\noindent
{\bf Definition-11 } The tensors of the form $ x \ot y$, where $ x\in X  $ and $ y\in Y $ are called elementary
tensors. They span the entire
tensor product space $ X \ot Y $. \\

\noindent
{\bf Caution : } The set of all elementary tensors is not a linearly independent set; for the reason that
there are too many of them. Hence, even though they span the entire vector space $X \ot Y, $ they do not
constitute a basis. One important consequence of this that representation of an arbitary tensor in terms of
elementary tensors is not unique. Two different looking tensors may actually turn out to be equal ! \\

\noindent
The elementary tensor of the form $ (x+x')\ot y $ acts on a bilinear form $f $ in the following way. 
$ [(x+x')\ot y](f)= f(x+x',y)=  f(x ,y) + f(x',y) = (x \ot y)(f) +  (x'\ot y ) (f) =[ (x \ot y) +  (x'\ot y
)](f) $. Since, this is valid for every bilinear form $f$, it follows  that,  $ (x+x')\ot y = x\ot y + x'\ot y
$. Similar reasoning leads to the following list of identities. 

\begin{enumerate}

{\item $( x_1+x_2 )\ot y = x_1 \ot y +x_2 \ot y $ }

{\item $ x \ot( y_1 + y_2 ) = x \ot y_1 +x \ot y_2  $}

{\item $ (\alpha x )\ot y = \alpha (x \ot y ) = x \ot (\alpha y )$ }

{\item $0_X \ot y = x \ot 0_Y = 0_{X \ot Y} $ }

\end{enumerate}

\noindent
where $ x, x_1,x_2  \in X $ ; $ y, y_1,y_2 \in  Y $ and $ \alpha  $ is a complex number. The symbols, $0_X ,
0_Y $ and $  0_{X \ot Y}$ denote the null vectors of the vector spaces $X , Y $ and $ X \ot Y $
respectively. \\

\noindent 
These properties are  summarised by saying that the tensor product $ \ot $, is a bilinear map from $ X \times
Y $ to $ X \ot Y$. Note, this map takes the pair $(x,y)$ to $ x \ot y $. From an abstract$^{9}$ point of view
this is the most important bilinear map for the pair of vector spaces ($X$ ,$Y$). If you call this bilinear
map $b$, then $ b: X \times Y \rightarrow X\ot Y $, and $b(x,y)=x\ot y$. Now given any vector space $W $ and a
bilinear map $f:X\times Y \rightarrow W $, there is a unique linear map $T_f: X \ot Y \rightarrow W$ such that
$f$ can be factored as, $f= T_f \circ b $. That is, $f(x,y)= T_f \circ b(x,y)= T_f( x\ot y)$,  $  x \in X $ ,
$ y \in Y  $. Essentially, the pair $(b, X \ot Y )$, -converts- bilinear maps on $ X\times Y $, into linear
maps on $  X \ot Y $.\\

\noindent Suppose, $x_1$ and $x_2$ are two linearly dependent vectors in $ X$ and $y_1,$ and $ y_2 $ are
arbitrary vectors  in  $ Y$, then the tensor of the form $ t= x_1\ot y_1 + x_2 \ot y_2 $ is actually an
elementary tensor. This is because, $ x_1\ot y_1 + x_2 \ot y_2 = (\alpha x) \ot y_1 + (\beta x) \ot y_2 = x
\ot (\alpha y_1) + x \ot (\beta y_2) = x \ot (\alpha y_1 + \beta y_2) = x\ot y $ where $ y=\alpha y_1 + \beta
y_2 $. Here, we have  made use of the fact that $\{x_1, x_2 \} $ is a linearly dependent set and hence $x_1=
\alpha x$ and $ x_2= \beta x $ for some $ x \in H_1$. The rest of the steps follow from the bilinear
properties of the tensor product listed above. Thus, a linear combination of elementary tensors is a
non-elementary tensor if and only if it cannot be reduced to an elementary tensor as we have just
demonstrated. An example of a non-elementary tensor is  $ \tau =u_1\ot v_1 + u_2 \ot v_2 $, where $ \{ u_1,u_2
\} $ is a linearly independent set in $H_1$ and $ \{v_1,v_2 \} $ is  a linearly independent set in $H_2$. Such
a $\tau $ can never be written in the form of $ x \ot y $. This important fact is also crucial for our final
result.\\

\noindent
{\bf Proposition - 8} Let $ X  \ot Y $ be the tensor product of vector spaces $X$ and $Y.$ Suppose, $ \{u_1,
u_2 \}  $ is a linearly independent subset of $X$ and $ \{ v_1,v_2 \} $ is a linearly independent subset of
$Y$.  Then a tensor of the form $  u_1\ot v_1 + u_2 \ot v_2 $ is not equal to $ u\ot v $ for any $ u \in X $
and $ v \in Y$. Hence, a non-elementary tensor can never be expressed as a scalar multiple of an elementary
tensor.\\ 

\noindent

{\bf Proof : } (cf. Appendix-C ) \\ 

\noindent
 With these tools we begin our study of  composite quantum states.\\ 

\subsection{ States of composite quantum systems }

\noindent
 Consider a composite quantum mechanical system that consist of  two particles, say, particle-1 and
particle-2. If the particle-1, as an individual entity, was modelled on a Hilbert space $H_1$ and the
particle-2, as an individual entity, was modelled on a Hilbert space $H_2$, then the composite system is
modelled on the tensor product space of $H_1 \ot H_2$. Hence, the composite states (both pure and mixed ) are
operators  that act on $H_1 \ot H_2$. First we shall look at pure states. Clearly, by definition-7, a pure
state of this composite  system is a self-adjoint, rank-one projection operator acting on $H_1 \ot H_2$. As
before, we shall denote it by $P_t$, where $ t \in H_1\ot H_2 .$ Observe, that $t$  could  either be an
elementary tensor or a non-elementary tensor. First we look at the case of elementary tensor.\\

\noindent
{\bf Proposition - 9 } A pure state $P_t :H_1 \ot H_2  \rightarrow H_1 \ot H_2  $, of a composite quantum
system on $H_1 \ot H_2 $,  where $t= x\ot y $, an elementary tensor is  a tensor product  of pure states of the
subsystems. Equivalently, $ P_{x\ot y}= P_x\ot P_y $. Here, $P_x$ and $P_y$ are the pure states of the
subsystems on $H_1$ and on $H_2$ respectively. \\ 

\noindent
{\bf Note } As such a state, is in the form of a (tensor) product of states of subsystems, it is called a
product state. Observe, this is a tensor product of operators. The set of all linear operators or matrices  on
a vector space itself is a vector space. Hence, tensor product of two such spaces of operators is well
defined. For example, if $M_2$ denotes the vector space of $ 2 \times 2 $ complex matrices,  then $M_2 \ot M_2
$ denotes the tensor product of $M_2$ with itself.\\ 

\noindent
{\bf Proof:}

\noindent
Let $ t=x \ot y,$ be  an elementary tensor in $H_1 \ot H_2$. Then $ P_t : H_1 \ot H_2 \rightarrow H_1 \ot
 H_2 $ acts on $ \tau \in H_1 \ot H_2 $  in the following way. $ P_t(\tau )= \la \la t , \tau \ra \ra t $.
 Here $ \la \la \tau, t \ra \ra $ denotes the inner product of the tensor product space $H_1 \ot H_2$. This 
 innerproduct is defined as \begin{equation*} \la \la u_1 \ot u_2, v_1\ot v_2 \ra \ra = \la u_1, v_1 \ra
 _{H_{1}} \times \la u_2, v_2 \ra _{H_{2}}  \end{equation*}

\noindent
 for elementary tensors and is extended to arbitrary tensors using the well known properties of inner
 product. In the following we shall suppress the subscripts  $H_i$,  on the inner products $\la  . , .
 \ra_{H_{i}} $   for the sake of readability. Let  $ \tau =  u \ot v $, then \\

$ P_t( \tau ) = P_{x \ot y} (u \ot v) $ \\
  
$ =  \la \la \, x\ot y ,    u \ot v \ra \ra [x \ot y ] $ (definition of $ P_{x \ot y} $ )\\

$ =  \la  x, u \ra \, \, \la y , v \ra \, \, [ x \ot y ] $ (definition of $ \la \la . , . \ra \ra )$ \\
  
$ = [\la  x, u \ra x \ot  \la y , v \ra   y]  $ (using the bilinearity of $\ot $ ) \\

$ = P_{x} u \ot P_{y} v  = [ P_x \ot P_y ]  (u \ot v ) $ \\

\noindent
Thus, $ P_{x\ot y} (u \ot v ) = [P_x \ot P_y ](u \ot v ) $ for an arbitrary elementary tensor $ (u \ot v ) $.
Since, $ P_{ x \ot y} $ is a linear operator, this equality  extends to non-elementary tensors  as well.
Thus,   $ P_{x\ot y} (\tau  ) = P_x \ot P_y (\tau  ) $,  for an arbitrary tensor $\tau \in H_1 \ot H_2 $.
Hence, $ P_{x\ot y}= P_x\ot P_y $. This proves proposition-9.\\

\subsubsection{ Separable states and entangled states }

\noindent When a composite system is in a product state $P_x \ot P_y $, one says that particle-1 is in the
state $P_x$ of the subsystem $H_1$ and particle-2 is in the state $P_y$ of the subsystem $H_2$. This implies,
that these two particles  act independent of each other. That is, there is no correlation between them. This
situation is analogous to the case in probability theory, where  two  random  variables $x$ and $y$ are said
to be independent if their composite probability density $ \phi (x,y) $, can be written as a product of
individual densities, say, as $ \phi(x,y) = \phi_1(x) \times \phi_2(y)$. In fact, not only a product state but
any convex combination of such product states also do not have a strong correlation between the  subsystems.
Such a state is called a separable state.\\

\noindent
{\bf Definition-12 } A state $\rho$ of a composite system $H_1 \ot H_2 $ is said to be a separable state if it
can be expressed as  $  \sum_{i=1}^m p_i (P_{x_{i}} \ot P_{y_{i}}) = \sum_{i=1}^m p_i \, P_{x_{i}
\ot y_{i}} $ , where $P_{x_{i}} $ and $P_{y_{i}} $ are the pure states of the subsystems $H_1$ and $H_2$
respectively for  $ 1 \le i \le m $. Here,  $ p_i \ge 0 $ for  $ 1 \le i \le m $ and $ \sum_{i=1}^m p_i =1 $.
Note,  $ P_{x_{i}} \ot P_{y_{i}} = P_{x_{i} \ot y_{i}} $ represents a pure state of the composite system
associated with the elementary tensor $ x_{i} \ot y_{i}$.  Observe, when  $m=1$, a separable state becomes a
product state. Thus, a separable state is a convex combination of product states, that is, states of the form
$ P_{x_{i}} \ot P_{y_{i}} $. \\

\noindent
{\bf Proposition-10}  A pure state $P_t$, of the composite system $ H_1 \ot H_2 $, where $t$ is an
elementary tensor, is a separable state.\\

\noindent
{\bf Proof : }
\noindent
This is because by Proposition-9, every pure  state $ P_t$, where $t$ as an elementary tensor is equal to $P_x
\ot P_y$, for some $x \in H_1 $ , $ y \in H_2$. Hence, such a state is a separable state.\\

\noindent 
Separable states are also called as classically correlated states$^{10}$. This is justified because,
as we saw in section 2.1, every classical composite state is in the form of a separable state. Right now it is
not at all obvious that there are  states  that are not separable.  One expects non-separable states to have
certain degree of correlation between its subsystems. A composite states that is not in the form of a
separable state is called an entangled state. \\ 

\noindent
{\bf Definition-13 } A composite state that is not  separable  is called an entangled state.\\

\noindent
Before we get to look at entangled states, we need one more result on separable states. This result is known
as the range criterion in quantum information theory.\\

\noindent 
{\bf Proposition -11 } The range of a separable state $\rho_s : H_1 \ot H_2 \rightarrow H_1 \ot H_2 $, which
is a subspace of  $ H_1 \ot H_2 $, is spanned by elementary tensors. That is,  the subspace range$\rho_s$, has
a basis that consisits entirely of elementary tensors. For example, if $\rho_s =p_1 P_{x_{1} \ot y_{1}} + p_2
P_{x_{2} \ot y_{2}} $, then range$(\rho_s)$=span $\{ x_{1} \ot y_{1} ,x_{2} \ot y_{2} \}$. \\

\noindent

{\bf Proof :}  ( cf. Appendix-D ) \\

\noindent

	There are plenty of pure states in a composite system, which are of the form $P_{\tau}$, where $\tau $
is a non-elementary tensor. This  is the case, for example, if  $  \tau =u_1\ot v_1 + u_2 \ot v_2 $, where $
\{ u_1,u_2 \} $ is a linearly independent set in $H_1$ and $ \{v_1,v_2 \} $ is  a linearly independent set in
$H_2$. \\

\noindent

Now we prove that every pure state that is associated with a non-elementary tensor is not a separable state. 

\noindent

{\bf Proposition-12 } A composite pure state $P_{\tau} $, a  rank one, self-adjoint, projection operator
acting on  $H_1 \ot H_2 $,  where $\tau $ is a non-elementary tensor in $H_1 \ot H_2 $ represents an entangled
state.
\noindent

{\bf Proof:} Assume the contrary. That is, let $P_{\tau}= \rho_s $, where $\rho_s$ is a separable state. Let
$  \tau=u_1\ot v_1 + u_2 \ot v_2 $, be the non-elementary tensor. Then  $ \{ u_1,u_2 \} $ is a linearly
independent set in $ H_1$ and $ \{v_1,v_2 \} $ is  a linearly independent set in $ H_2$. It follows, that the
range of $P_{\tau}$ is equal to the range of $  \rho_s $. $P_{\tau} $, being a pure state has a one
dimensional range spanned by $\tau $. That is, the range of $P_{\tau }$ is the set  $ \{ \alpha \tau =\alpha
(u_1\ot v_1 + u_2 \ot v_2) : \alpha \in C \}$, a one dimensional subspace of $H_1\ot H_2 $. By proposition- 11
, the range of a separable state $ \rho_s$ is spanned by elementary tensors. Since, $P_{\tau}= \rho_s $, the
range of $ \rho_s $ is also a one dimensional subspace spanned by an elementary tensor, say, $ x \ot y $. Thus, the
 Range$ ( P_{\tau}) =  \{ \alpha \tau =\alpha (u_1\ot v_1 + u_2 \ot v_2) : \alpha \in C
\}$ = Range ($ \rho_s$ )= span $\{ x \ot y   \}$   = $ \{\beta ( x \ot y) :  \beta \in \mathbb{C} \} $. By
Proposition- 8,  it is not possible to express a non-elementary tensor as a scalar multiple of an elementary tensor. Thus, we
have reached a contradiction. Hence, $P_{\tau} \ne \rho_s $, for any separable $ \rho_s $. So we conclude that
$P_{\tau}$, when $\tau $ is a non-elementary tensor is an entangled state.\\

\noindent

In the case of  classical states,  every composite pure state turned out to be a product of pure states of
subsystems, called a product state   and hence a non-entangled state. Moreover, as every state is a convex
combination of pure states, all states turn out to be convex combination of such product states, that is,
non-entangled states. However, as we have realised, the pure states of composite quantum systems that are
associated with non-elementary tensors are  entangled. In fact, there are also mixed states  which are
entangled in the case of  quantum mechanics. In contrast, no classical state, either pure or mixed  is an
entangled state.\\
\noindent

{\bf Acknowledgement} The author thanks the members of the theory group at MSD, IGCAR for their active
participation in many  discussions on entanglement and the organisers of the conference, " Entanglement in
quantum condensed matter," held at the Institute of Mathematical sciences, Chennai,  during 17-29, November,
2008.\\

\section{Appendix}

\begin{center}

{ \bf Appendix-A }

\end{center}

\noindent
{\bf Proposition-4} If $H$  is a complex vector space then $T: H \rightarrow H $ is a self-adjoint operator if
and only if $ \la T u, u \ra = \la u, T u \ra  $  for every $u \in H$.\\ 

\noindent

{\bf Proof: } Let  $H$ be a complex vector space. We have to show that $ \la T u, u \ra = \la u, T u \ra  $ 
for every $u \in H$ is equivalent to $ \la T u, v \ra = \la u, T v \ra $ for every $u,v \in H$.  Suppose, $
\la T u, v \ra = \la u, T v \ra  $,  for every $ u, v \in H$ then it is obvious by putting $u=v$ that $ \la T
u, u\ra = \la u, T u \ra = \overline{\la T u, u\ra} $ for every  $ u \in H $.   In the other direction,
suppose $ \la T x, x \ra = \la x, T x \ra $ for every $ x \in H $  then $ \la T (u+\alpha v), (u+ \alpha v
)\ra = \la (u+ \alpha v), T (u+ \alpha v )\ra $ for every $u,v \in H $ and $ \alpha \in \mathbb{C} $.
Expanding the above expression leads to the equality  $\la  u, T \alpha v \ra + \la \alpha v , T u \ra $ = 
$\la T u, \alpha v \ra + \la T \alpha v ,  u \ra $. Which implies  Im ($ \alpha  \la u, T v \ra $ )  = Im ( $
\alpha \la T u,  v \ra $ ). We use Im(z) and  Re(z) to  denote the imaginary and real part of complex number
$z$ respectively. The equality being valid for every complex number $\alpha $; Choosing $\alpha =i$, where
$i^2 =-1, $ it follows  Re$ (\la u,Tv \ra ) $= Re $ \la Tu, v \ra $  and choosing  $ \alpha =1 $, it follows 
Im $( \la u,Tv \ra ) $= Im $ (\la Tu, v \ra ) $. Thus  $  \la u,Tv \ra  =   \la Tu, v \ra  $.\\

\noindent
{\bf Remark: } Proposition-4 cannot extended to real vector spaces. For example, the $2 \times 2 $ real
matrix $A$, with $ A_{1,1}=A_{2,2}=1 , A_{1,2}=2 $ and $ A_{2,1} =0,$ considered as  an operator acting on
$R^2$ is not self-adjoint, even though $ \la x, A x \ra $ is a real number for every $x \in R^2 $.\\

\begin{center}

{ \bf Appendix-B }

\end{center}

\noindent
{\bf Proposition-5 } In a complex vector space $H$, the following two statements about a linear operator $T: H
\rightarrow H $ are equivalent.\\  

\noindent
1) $T = B^* B $ for some operator $B$. \\  
2) $ \la x, T x \ra \ge 0 $ for every $x \in H$.\\

\noindent
If $T =B^*B$, then $ \la x, T x \ra = \la x, B^*B x \ra =\la Bx, B x \ra = ||Bx||^2 \ge 0$, by the axioms of
norm. In the other direction, if $ \la x, T x \ra  \ge 0 $ for every $x \in H$ then by proposition-4 it
follows  that $T$ is self-adjoint. We claim that all the eigen values of $T$ are non-negative. Suppose $u$ is
an eigenvector of $T$, with eigenvalue $\lambda $, then $ \la u, T u \ra  = \la u, \lambda u \ra  = \lambda
\la u,  u \ra \ge 0 $, which implies $\lambda $ and hence  all the eigenvalues of $T$ are positive. Since $T$
being self-adjoint the eigenvectors of $T$ form a  basis of $H$. Then, such a $T$ can be expressed, in the
basis consisting of its eigenvectors, as a diagonal matrix with its  non-negative eigenvalues $ \lambda_i $ as
diagonal elements. By  a diagonal matrix we mean a matrix whose non-diagonal entries are all zero.  We denote
the matrix that represents the operator $T$ as $[T]$.  Thus we have $[T]$ = diag$ (\lambda_1,
\lambda_2,....,\lambda_n)$ where $ \lambda_i \ge 0 $ for $ 1 \le i \le n $. Now one can write  $[T]=[B^*][
B]$, where $[B] $= diag $( \sqrt{\lambda_1}, \sqrt{\lambda_2},... ,\sqrt{\lambda_n}) $. This completes the
proof.\\

\begin{center}

{ \bf Appendix-C }

\end{center}

\noindent
{\bf Proposition-8 } Let $ X  \ot Y $ be the tensor product of vector spaces $X$ and $Y$. Suppose, $ \{u_1, u_2
\}  $ is a linearly independent subset of $X$ and $ \{ v_1,v_2 \} $ is a linearly independent subset of $Y$.
Then a tensor of the form $  u_1\ot v_1 + u_2 \ot v_2 $ is not equal to $ u\ot v $ for any $ u \in X $ and $ v
\in Y. $\\

\noindent
{\bf Proof :}  
We have to show that $ u_1\ot v_1 + u_2 \ot v_2 \ne  u\ot v $ for any $ u \ot v \in X \ot Y.$ Equivalently, $
u_1\ot v_1 + u_2 \ot v_2 -  u \ot v \ne 0 $  for any $ u \ot v \in X \ot Y.$ We assume the contrary and reach
a  contradiction. Let   $ u_1\ot v_1 + u_2 \ot v_2 -  u \ot v = 0 $.  Recall, if a tensor $\tau \in X \ot Y $
is zero  then $ \tau (f)=0 $, for every bilinear form $f: X\times Y \rightarrow \mathbb{C} $. Specifically, $
\tau (f)$ is defined  such that, if $ \tau = u\ot v $, then $\tau (f) $ =  $ (u\ot v )(f)= f(u,v ) $. If $ 
\tau = u_1\ot v_1 + u_2 \ot v_2 $, then $ \tau (f) = (u_1\ot v_1 + u_2 \ot v_2) (f)= (u_1\ot v_1)(f) + (u_2
\ot v_2)(f)= f(u_1,v_1)+ f(u_2, v_2) $, for every bilinear form $f$. So if $ (u_1\ot v_1 + u_2 \ot v_2 -  u\ot
v )$  is a zero tensor then $ ( u_1\ot v_1 + u_2 \ot v_2 -  u \ot v ) (f)= 0 $ for every bilinear form  $f$. 
Note, $( u_1\ot v_1 + u_2\ot v_2 -  u \ot v ) (f)= ( u_1\ot v_1)(f) + (u_2\ot v_2)(f)  -  (u \ot v )(f) = f
(u_1\ot v_1)  + f(u_2 \ot v_2) - f( u \ot v ) = f(u_1, v_1) + f(u_2 , v_2) -  f (u , v ) $, for every bilinear
form $f$.\\

\noindent  Now we construct some bilinear forms, using the linear functionals that act on $X$ and $Y$. We
shall use the symbol $\phi $ and $ \psi $ for an arbitrary linear functional  in $ X^*$ and $ Y^* $
respectively. Observe, if  $\phi \in X^* $ and $ \psi  \in Y^*$, then $ \phi : X \rightarrow \mathbb{C} $ and
$ \psi : Y \rightarrow  \mathbb{C}  $.  Then we define a bilinear form $ \phi \times \psi $, on   $ X \times
Y$ such that   $ (\phi \times \psi)(x,y)= \phi(x) \times \psi(y)$, $ x \in X $, $ y \in Y $.  Since, $ \{ u_1,
u_2 \} $ is a linearly independent set in $X$, we can construct an ordered  basis of $X$, which includes $ u_1
$ and $ u_2 $ as its first two elements. That is,  $ \{ u_1 , u_2, u_3, ..., u_n \} $ is a basis of $X$. Then,
let $ \phi_1 : X \rightarrow \mathbb{C}  $ be a linear functional  in the dual space $ X^* $, such that $
\phi_1(u_1) = 1 $ and   $ \phi_1(u_k) = 0 $ for all  $ k \ne 1 $. Such a linear functional   always exist as
we discussed above in section 4.1 on dual spaces. Similarly, as $ \{ v_1,  v_2 \} $ is a linearly independent
set in $Y$, one can construct an ordered  basis of $Y$, which includes $ v_1 $ and $ v_2 $ as its first two
elements. That is, $ \{ v_1 , v_2, v_3, ..., v_m \} $ is a basis of $Y $. Then, let $ \psi_2 : Y \rightarrow
\mathbb{C}  $ be an element in the dual space $ Y^* $, such that $ \psi_2(v_2) = 1 $ and   $ \psi_2(v_k) = 0 $
for all  $ k \ne 2$.\\

 \noindent
{\bf Step-1 } We claim that $v_1 $ and $ v$ are linearly dependent. Consider, a bilinear form $f$, such that
$ f(x,y)= \phi_1(x) \times \psi(y)  $, $ x \in X $, $ y \in Y $, where $\phi_1 $ is the particular linear
functional as defined above and $\psi$ is an arbitrary linear functional in $Y^*$. We shall denote this
bilinear form as $ \phi_1 \times \psi $. Since, $( u_1\ot v_1 + u_2 \ot v_2 -  u \ot v ) (f) = 0$, for every
bilinear form $f$, it follows $( u_1\ot v_1 + u_2 \ot v_2 -  u \ot v ) ( \phi_1 \times \psi)= 0 $. Which
implies $ ( u_1\ot v_1)( \phi_1 \times \psi) + ( u_2 \ot v_2) ( \phi_1 \times \psi) -  ( u \ot v ) ( \phi_1
\times \psi)= \phi_1( u_1) \psi (v_1) +  \phi_1(u_2) \psi( v_2) -  \phi_1(u) \psi ( v ) = 0 $. By the 
definition of  $\phi_1$, $\phi_1(u_1)=1 $ and  $ \phi_1(u_2)=0 $. Hence, we have $ \psi(v_1) - \phi_1(u)
\psi(v)=  \psi ( v_1 - \phi_1(u) v) = 0 $. Since, $\psi $ is  an arbitrary linear functional  in $Y^*$, it
follows $ v_1 - \phi_1(u) v = 0 $. Here, we are using the fact (cf. Section 4.1) that if $ \psi(y) = 0 $ for
every $ \psi \in Y^* $ then $y=0 $. Since, $ v_1 - \phi_1(u) v = 0 $, we conclude that $ v_1 $ and $ v$ are
linearly dependent. Hence, $ \psi_2(v)= \psi_2 (\alpha v_1)=\alpha \psi_2(v_1)=0 $, as by definition $
\psi_2(v_1)=0 $. 

\noindent
{\bf Step-2 } Now we claim $u_2$ is zero, which is in contradiction to the fact that $\{ u_1, u_2 \} $ is
linearly independent. Recall, any set  that contains a null vector is linearly dependent. Consider a  bilinear
form $f$, such that $f= \phi \times \psi_2 $, where $\phi $ is an arbitrary linear functional in $X^*$ and
$\psi_2 $ is the specific linear functional in $ Y^*$, that was defined above. Then, $ f(x,y)= \phi(x) 
\psi_2(y)$, $ x \in X $, $ y \in Y $. Since, $( u_1\ot v_1 + u_2 \ot v_2 -  u \ot v ) (f) = 0$, for every
bilinear form $f$, it follows  $ ( u_1\ot v_1 + u_2 \ot v_2 -  u \ot v ) ( \phi \times \psi_2)=  ( u_1\ot
v_1)( \phi \times \psi_2)  + (u_2 \ot v_2) ( \phi \times \psi_2) -  (u \ot v ) ( \phi \times \psi_2) =0 $.
Which implies, $ \phi(u_1) \psi_2(v_1) + \phi(u_2) \psi_2(v_2) - \phi(u) \psi_2(v)= \phi(u_2) =0 $. Here we
have used the properties of $ \psi_2$  that $ \psi_2 (v_2)= 1$, $ \psi_2 (v_1)=0 $, and the fact $\psi_2(v)= 0
$, which was obtained at the end of step-1. Since, $ \phi(u_2)= 0 $ and $ \phi $ is an arbitrary linear
functional it follows that $u_2=0 $. This is a contradiction, because the set $\{ u_1, u_2 \} $ was by
assumption a linearly independent set and hence cannot contain a null vector. Hence, we conclude that $ u_1\ot
v_1 + u_2 \ot v_2 \ne  u\ot v $ for any $ u \ot v \in X \ot Y $. \\

\begin{center}

{ \bf Appendix-D }

\end{center}

\noindent
{\bf Proposition -11  } The range of a separable state $\rho_s : H_1 \ot H_2 \rightarrow H_1 \ot H_2 $, which
is a subspace of  $ H_1 \ot H_2 $, is spanned by elementary tensors. That is, the subspace range($ \rho_s $)
has a basis that consists entirely of elementary tensors. For example, if $\rho_s =p_1 P_{x_{1} \ot y_{1}} +
p_2 P_{x_{2} \ot y_{2}} $, then range$(\rho_s)$=span $\{ x_{1} \ot y_{1} ,x_{2} \ot y_{2} \}$. 

\noindent
{\bf Proof: } Recall, that the range of a linear operator $T: H \rightarrow H $, is the set $\{ T(x) : x \in H
\}$; which is a subspace of $H$. Note,  $ P_{x \ot y} (t) = \la x \ot y, t \ra (x \ot y), $   $  t \in H_1 \ot
H_2 $. \\

\noindent
Let $  \rho_s =  p_1 P_{x_{1} \ot y_{1}} + p_2 P_{x_{2} \ot y_{2}} $, where  $ p_1, p_2 \ge 0 $ and $ p_1
+p_2= 1$. We split the proof into two parts.\\ 

\noindent
{\bf Case(i):} Assume the  set $\{ x_{1} \ot y_{1} ,x_{2} \ot y_{2} \} $ to be  linearly dependent. Then,  $ x_{1}
\ot y_{1}= \alpha  (x \ot y) $ and  $ x_{2} \ot y_{2} =\beta   (x \ot y )$ for some $  \alpha , \beta \in
\mathbb{C}$ , $  x \ot y \in H_1 \ot H_2 $. Thus, $ \rho_s (t)= p_1 \la x_1 \ot y_1, t \ra (x_1 \ot y_1), +
p_2\la x_2 \ot y_2, t \ra (x_2 \ot y_2 )= p_1\overline{\alpha} \alpha \la x \ot y, t \ra (x \ot y) + 
p_2\overline{\beta} \beta \la x \ot y, t \ra ( x \ot y) = \gamma  (x \ot y) $, \, $ \gamma \in \mathbb{C}$ and
$ t \in H_1 \ot H_2 $. Hence, the range of $\rho_s $ is the span of the particular element $ x \ot y $, an
elementary tensor.\\

\noindent
{\bf Case (ii): } Assume the set $\{ x_{1} \ot y_{1} ,x_{2} \ot y_{2} \}  $ to be linearly independent. Now,
we claim that there is a $u_0 \in H_1 \ot H_2 $, such that $P_{x_1\ot y_1}(u_0)= 0 $ and $  P_{x_2\ot
y_2}(u_0) \ne 0 $. In that case, $ \rho_s(u_0)= p_2 P_{x_2\ot y_2}(u_0) = p_2 \la x_2\ot y_2, u_0 \ra (x_2\ot
y_2 )= \alpha (x_2\ot y_2 )\ne 0,$  and hence $(x_2 \ot y_2 )$ is in the range of $\rho_s$. Suppose the
contrary, that is, assume that there is no  such $u_0$. This would mean, for any $u$, for which $P_{x_1\ot
y_1}(u)=0 $ it follows $ P_{x_2\ot y_2}(u) = 0 $, as well. Recall, that  the operator $ P_{x \ot y}$, takes
every vector that is orthogonal to $x \ot y$ to null vector.  We denote the set of all vectors that are
orthogonal to $x \ot y $ by $ (x \ot y)^{\perp} $. Note, if dim($H_1 \ot H_2$) = n, then $  (x \ot y)^{\perp}
= \{ t \in H_1\ot H_2 : \la (x \ot y), t \ra =0 \}$ is a $(n-1)$ dimensional subspace of  $ H_1 \ot H_2 $.
Since, $P_{x_1\ot y_1}(u)=0 $  implies $ P_{x_2\ot y_2}(u) = 0 $,  we have $ (x_1 \ot y_1)^{\perp} \subset
(x_2 \ot y_2)^{\perp} $. Observe, dim$ (x_1 \ot y_1)^{\perp}$= n-1 =  dim $ (x_2 \ot y_2)^{\perp} $, which
implies $  (x_1 \ot y_1)^{\perp} =(x_2 \ot y_2)^{\perp} $. Note,  since $ \{ (x_1 \ot y_1)^{\perp} \}^{\perp}
$ = span $\{ x_1 \ot y_1 \} $, and $ ((x_1 \ot y_1)^{\perp})^{\perp} = ((x_2 \ot y_2)^{\perp})^{\perp} $ one 
concludes that span $\{ x_1 \ot y_1 \} $ = span $\{ x_2 \ot y_2 \} $. This implies that $ x_{1} \ot y_{1}$ and
$ x_{2} \ot y_{2}$ are linearly dependent. This  is a contradiction. Thus, our claim, that there is a $u_0 \in
H_1 \ot H_2 $, such that $P_{x_1\ot y_1}(u_0)= 0 $ and $  P_{x_2\ot y_2}(u_0) \ne 0 $  is true and hence $(x_2
\ot y_2 )$ is in the range of $\rho_s$. Reversing the role of $ x_1 \ot y_1 $ with that of  $  x_2 \ot y_2  $,
one concludes that  $(x_1 \ot y_1 )$ is also in the range of $\rho_s$. Thus, it is clear that the range of
$\rho_s$ is spanned by the elementary tensors $x_{1} \ot y_{1} $, and $ x_{2} \ot y_{2}$. This proves the
proposition. \\

\begin{center}

{ \bf Appendix-E }

\end{center}

\noindent {\bf Dirac's notation :} Let X be a vector space with an innerproduct denoted as $\la ., . \ra $ and
$X^*$ its dual as defined in section 4.1. In Dirac's notation, $ x \in X $ is written as $ | x \ra $, and is
called a ket vector and $ \phi \in X^* $ is written as $ \la \phi | $, and is called a  bra vector. Similarly,
what is written as $ \phi(x) $, in our notation, where $ \phi \in X^* $ and $ x \in X$ is written as $ \la
\phi | x \ra $ in Dirac's notation. Right now, the symbol $\la  . | . \ra $ that occurs in Dirac's notation  $
\la \phi | x \ra $ cannot be interpreted as an innerproduct. This is because the the expression -$ \la \phi, x
\ra  $ - does not make sense as $ \phi \in X^* $ and $ x \in X$, live in  distinct vector spaces. However,
Reisz representation theorem$^{11}$ says that every continuous linear functional $ \phi \in X^*$ can be
represented as $ \phi(x)=\la v_{\phi}, x \ra \, , x \in X$,  where $v_{\phi}  \in  X$ is fixed unique vector
associated with $\phi $. This correspondence, $ \phi \in X^* \rightarrow  v_{\phi}\in X $, is a linear map
that establishes a one to one correspondence between $ X^* $ and $X$. The linearity of this correspondence
ensures that if $ \phi_{1} $ and $ \phi_{2} $ are independently mapped to $ v_1 $ and $v_2 $ respectively
then  $\phi_{1} + \phi_{2} $ gets mapped to $ v_1 + v_2 $. On the other hand, as we saw in section 4.1, every
$ v \in X $ gets associated with a linear functional $ \phi_v$, where $ \phi_v(x) = \la v,x \ra \, ,\, x \in X
$.  Thus we have a  natural ( independent of basis) means  of identifying elements of $X^* $ with that of $
X$. In other words, this allows us  to treat the $ \phi \in X^* $ as if it were  $ v_{\phi} \in X  $ in the 
sense; $ \la \phi | x \ra = \phi (x) =  \la v_{\phi}, x \ra $, where the last equality  makes  use of the
Reisz representation theorem. \\

\vspace{.2in}

{\bf References } \\

$^{1} $ Asher Peres, " Quantum Theory : Concepts and Methods, Kluwer Academic Publishers,Chapter-5, 1993.(cf.
Chapter-5, Composite systems, Chapter-6, Bell's theorem )\\

$^{2}$P.K.Aravind," Quantum mysteries revisited again," Am. J. Phys.{\bf 72} (10)1303-1306(2004)\\

$^{3}$ R.F.Werner," Quantum information theory- An invitation", In G.Alber, T.Beth, M.Horodecki, R.Horodecki,
M.Rotteler, H.Weinfurter,R.F.Werner and A.Zeilinger, "  Quantum information: An introduction to basic
theoretical concepts and experiments( Springer Tracts in Modern Physics, {\bf 173}; Springer-verlag,2001).\\

$^{4}$ Erling Stormer," Extension of positive maps into $B(H)$  ", Journal of functional analysis, {\bf 66 },
235-254 (1986). Lemma-2.2, (page-237) is a generalised version of  our Proposition-6. Operator theorists,
define a state as a positive linear functionals from the space of operators to complex numbers, that takes the
identity operator to the complex number 1. Here, positivity means, the linear functional takes positive
operators to positive real numbers. This description is a generalisation of density operators. In the abstract
setting of $C^* $-algebra, classical mechanical states are represented as elements of abelian algebra, while
quantum states are from a non-abelian algebra. \\

$^{5}$ A.Hobson, Concepts in Statistical Mehanics, Gordon and Breach, 1971, Newyork. (cf. Chapter-3, p-53 ;
Chapter-4,p-93 to p-100)\\

$^{6}$ The claim is that it is not possible to represent both position $ x $ and momentum $  p $ as finite
dimensional operators or matrices such that their commutator $ x p - p x  = ihI $. Here $I$ denotes the $ n
\times n $ identity matrix, $i^2 =-1 $  and $h$ is the planck's constant. If it were true, then taking trace
on both sides one gets  tr$(x p - p x )$= tr$( xp )$ - tr$( px)$= 0 = $inh $, a contradiction. Recall, tr $(A
+ B )$ = tr $ A $ + tr $ B $ and tr$(AB)$=tr$(BA)$.\\

$^{7} $ when a linear operator $T$, is represented as a matrix, $M_T$, then $Tr(T)=Tr(M_T)= $ sum of the
diagonal elements of $M_T$. Let dim$H$=2 and   $ B=\{e_1=(1,0)^{trp} , e_2=(0,1)^{trp} \}$, be the standard
orthonormal basis of $H$, where $trp$ denotes the transpose. Then, $Tr(M_T)=  \la e_1, M_T e_1 \ra +\la e_2,
M_T e_2 \ra = ({M_T})_{1,1} + ({M_T})_{2,2}$, is seen to be the sum of diagonal elements of $M_T$.\\

$^{8}$ Let, $ T : H \rightarrow H $. Our aim is to show, if $y_1$ and $y_2$  are in range(T),  then so is $
\alpha y_1 + \beta y_2 $ for every $ \alpha , \beta \in C $. Since $y_1$ and $y_2$  are in the set range(T),
there are vectors $x_1$ and $x_2$ in $H$, such that $T(x_1)=y_1$ and $ T(x_2)=y_2 $. As $T$ is a linear
operator it follows $T( \alpha x_1 + \beta x_2 ) = \alpha T(x_1) + \beta T(x_2) = \alpha y_1 + \beta y_2 $.
Thus, if $y_1$ and $y_2$ are in the range of  $T$, then so is every linear combination of $y_1$ and $y_2$.
Hence,  range(T) is  a subspace of $H$. \\

$^{9} $ Raymond A. Ryan,  "Introduction to tensor products of Banach spaces, " Springer-verlag, 2002.
(Chapter-1).\\

$^{10}$ R,F.Werner, " Quantum states with Eienstein-Podolsky-Rosen correlations admitting a hidden variable
model," Physical Review A. {\bf 40}, 8, 4277-4281 (1989).\\

$^{11}$ Martin Schecter, " Principles of functional analysis, ", Academic press.(1971)

\end{document}